\documentclass[aps,prb,twocolumn,groupedaddress,showpacs]{revtex4}

\usepackage{graphicx,amsmath}

\begin{document}

\title{Entangled photon pairs from a quantum dot cascade decay: \\
the effect of time-reordering}

\author{Filippo Troiani}
\affiliation{CNR-INFM National Research Center S3  c/o Dipartimento
di Fisica via G. Campi 213/A, 41100, Modena, Italy}

\author{Carlos Tejedor}
\affiliation{Departamento de F\'{\i}sica Te\'orica de la Materia
Condensada, Universidad  Aut\'onoma de Madrid, Cantoblanco 28049
Madrid, Spain}

\begin{abstract}
Coulomb interactions between confined carriers remove degeneracies in
the excitation spectra of quantum dots. This provides a {\it which path
information} in the cascade decay of biexcitons, thus spoiling the
energy-polarization entanglement of the emitted photon pairs.
We theoretically analyze a strategy of color coincidence {\it across
generation} (AG), recently proposed as an alternative to the previous,
{\it within generation} (WG) approach. We simulate the system dynamics and
compute the correlation functions within the density-matrix formalism.
This allows to estimate quantities that are accessible by a
polarization-tomography experiment, and that enter the expression of the
two-photon concurrence. We identify the optimum parameters within the AG
approach, and the corresponding maximum values of the concurrence.
\end{abstract}


\maketitle

\section{Introduction}

\begin{figure}
 \begin{center}
 \includegraphics[width=0.8\columnwidth]{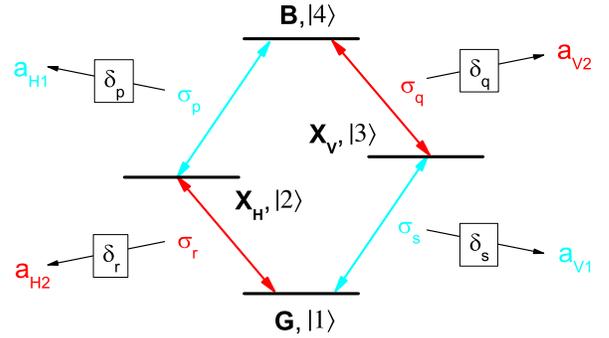}
 \caption{(Color online) Level structure of the quantum dot.
 The fine-structure splitting and the biexciton binding energy are
 given respectively by
 $ \delta_{HV} \!\! = \!\! E_3\!-\!E_2 $
 and
 $ \Delta_B \!\!=\!\! E_2 + E_3 - E_4 $.
 In the WG strategy, the ideal case corresponds to the biexciton and
 exciton emission frequencies being independent on the polarization:
 $ \omega_1\!\equiv\!\omega_p \! = \!\omega_q $ and
 $ \omega_2\!\equiv\!\omega_r \! = \!\omega_s $ ($\delta_{HV}\! = \! 0$).
 In the AG approach the photon emitted in the
 $B \rightarrow X_H$ ($B \rightarrow X_V$) transition matches the color of
 that emitted by the $X_V \rightarrow G$ ($X_H \rightarrow G$) decay
 ($\Delta_B \!=\! 0$).
 Therefore, $ \omega_1\!\equiv\!\omega_p \! = \!\omega_s $ (cyan arrows) and
            $ \omega_2\!\equiv\!\omega_q \! = \!\omega_r $ (red arrows).
 In order to erase the which-path information provided by the emission
 order of the $\omega_1$ and $\omega_2$ photons, which is opposite in
 the two paths, the
 photons are delayed by $\delta_\chi$ ($\chi = p,q,r,s$).}
 \label{fig1}
 \end{center}
 \end{figure}

In view of their peculiar level structure, semiconductor quantum dots
(QDs) are considered promising sources {\it on demand} of
entangled photon pairs.~\cite{benson,stace}
Although alternative strategies have been envisaged, based on the use
of single-photon sources and postselection,~\cite{fattal04_3,benyoucef04}
the possibility of deterministically generating frequency-polarization
entangled photon pairs by a single cascade emission has recently received
strong experimental support.~\cite{akopian,stevenson06,young06}
There, the radiative relaxation of the dot from the lowest biexciton
level generates a two-photon quantum state:
$ |\psi_{ph}\rangle = ( | \phi_H ; H , H \rangle +
                        | \phi_V ; V , V \rangle )/\sqrt{2}$,
where $H$ and $V$ are the two linear polarizations, whereas
$ | \phi_H \rangle $ and $ | \phi_V \rangle $ refer to the spectral
degrees of freedom.
In particular, $ | \phi_H \rangle $ ($ | \phi_V \rangle $) is the
wavepacket resulting from the sequential emission of two photons, with
central frequencies $\omega_p$ and $\omega_r$ ($\omega_q$ and $\omega_s$)
(Fig. \ref{fig1}).
Ideally,
$\omega_1\! \equiv \!\omega_p\! = \!\omega_q $
and
$\omega_2\! \equiv \!\omega_r\! = \!\omega_s $
(analogous equations hold for the relaxation rates):
therefore,
$ |\psi_{ph}\rangle \! = \! ( | H , H \rangle +
                        | V , V \rangle )/\sqrt{2}
\otimes |\phi\rangle $,
with $ | \phi \rangle \!\equiv\! | \phi_H \rangle \! = \! | \phi_V \rangle $.
In realistic conditions, however, the degree of entanglement is limited by
three main factors.~\cite{troiani06b,hohen07,hudson07}
First, an imperfect system excitation results in a finite probability
that the system does not undergo a single cascade decay, thus emitting
more (or less) than the two desired photons.
Second, the coupling of the confined excitons with phonons
tends to induce a loss of phase coherence in the state of the emitted
photons.
Third, the presence of an excitonic fine-structure splitting tends to make
photons emitted with orthogonal polarizations distinguishable in the
spectral domain:
$ \delta_{HV} \!= \!\omega_p \!-\! \omega_q\! = \!\omega_s\! - \!\omega_r\!
 \neq\! 0 $,
and therefore
$ | \langle\phi_H | \phi_V \rangle | \neq 1$.
This provides a which path information, which impedes to rotate the $H$ and
$V$ components of $ | \psi_{ph} \rangle $ one into another by linear optics
elements and to observe interference effects between them.
In optimizing the entangled-photon source, part of the recent effort
has been concentrated on the solution of the latter problem.
Possible strategies include quenching of the excitonic fine-structure
splitting $\delta_{HV}$
by means of magnetic field~\cite{stevenson06,young06}
or ac-Stark effect~\cite{jundt}, and
spectral filtering of the emitted photons.~\cite{akopian}
An ingenious alternative consists in engineering the system so as to
obtain color coincidence across generation (AG), rather than
within generation (WG).~\cite{finley,reimer,avron}
There, the QD spectrum is tuned in such a way to have vanishing
biexciton binding energy ($ \Delta_{B} = E_2 + E_3 - E_4 $),
so that the four
emission frequencies $ \omega_\chi $ ($ \chi \!\!=\!\! p,q,r,s $) reduce to
two:
$ \omega_1\!\equiv\!\omega_p \!=\! \omega_s $ (cyan arrows in Fig.~\ref{fig1})
and
$ \omega_2\!\equiv\!\omega_q \!=\! \omega_r $ (red arrows).
In the AG scheme, however, the which-path information is provided by
the order in which the $\omega_1$ and $\omega_2$ photons are emitted,
which is now opposite for the two polarizations.
In order to make the $H$ and $V$ paths spectrally indistinguishable,
the photons emitted in the
four modes $ \chi $ should be spatially separated and time delayed by
$\delta_\chi$. The time reordering implements a unitary transformation
($ U = U_H \otimes |H,H\rangle\langle H,H| + U_V \otimes |V,V\rangle\langle V,V| $ )
of the two-photons state, such
that:
$ | \langle \phi_H |U_H^\dagger U_V | \phi_V \rangle | > | \langle \phi_H |\phi_V \rangle |
$.~\cite{avron}

Hereafter, we investigate the viability and the limits of the AG
approach. More specifically, we verify to which extent the which
path information can be erased by introducing these frequency and
polarization selective delays. To this aim, we derive analytic
expressions for an entanglement measure (namely, the concurrence, \cite{wootters}
$\mathcal{C}$) of the two-photon state, and derive the expressions of the
delays $\delta^M_\chi$ that maximize $C$, as a function of the
emission rates $\Gamma_\chi$. This is the same as optimizing the
unitary quantum erasure $U$ of the which-path information, for a
given source. In addition, we maximize $C (\delta^M_\chi)$ with
respect to $\Gamma_\chi$, thus providing indications for the
optimization of the two-photon source. In semiconductor quantum
dots, the (relative) values of the exciton and biexciton relaxation
rates can only be engineered within a limited range of
values.\cite{wimmer} However, such ranges can be potentially
extended by coupling the QD to an optical microcavity. In the
weak-coupling regime, the effect of the cavity on the dot dynamics
essentially consists in enhancing the photon-emission rates of
resonant transitions (Purcell effect). Therefore, and in order to
allow analytic solutions, we don't include the degrees of freedom of
the cavity explicitly, but rather mimic its effect by enhancing
$\Gamma_\chi$.
We also neglect the effect of pure dephasing and imperfect initialization
of the QD state ({\it i.e.}, of realistic excitation conditions).
In fact, the way in which these affect the degree of frequency-polarization
entanglement is independent on the approach, AG or WG. Detailed discussions
on these effects, can be found in the
literature.~\cite{troiani06b,hohen07,hudson07}

The paper is organized as follows. In Sec. \ref{method} we introduce the
density matrix approach we use, and the correlation functions that enter
the calculation of the two-photon concurrence. Further details on the
method are given in the Appendixes. In Sec. \ref{results} we give the
expressions of the concurrence, and maximize it with respect to the
relevant parameters. In Sec. \ref{conclusions} we draw our conclusions.

\section{Method}\label{method}

The time evolution of the dot density matrix, $ \rho_{QD} $, is described
by the master equation
($\hbar\!=\!1$):
$ \dot\rho_{QD} = i[ \rho_{QD} , H ] + \sum_\chi\mathcal{L}_\chi \rho_{QD} $,
where
$ H = \sum_{k=1}^4 E_k | k \rangle\langle k | $
and
$ | k \rangle $
are the QD eigenstates (in the following, we take $E_1\!=\!0$).
The radiative relaxation processes are accounted for by the four
superoperators in the Lindblad form:
$ \mathcal{L}_\chi \rho_{QD} =
\sigma_\chi \rho_{QD} \sigma_\chi^\dagger -
(\sigma_\chi^\dagger\sigma_\chi\rho_{QD} +
 \rho_{QD}\sigma_\chi^\dagger\sigma_\chi )/2$,
with $ \chi = p,q,r,s $. Each of the ladder operators $ \sigma_\chi
$ corresponds to one of the optical transitions in the four-level
system: $ \sigma_p \!\equiv\! \sqrt{\Gamma_p}\, | 2 \rangle\langle 4
| $, $ \sigma_q \!\equiv\! \sqrt{\Gamma_q}\, | 3 \rangle\langle 4 |
$, $ \sigma_r \!\equiv\! \sqrt{\Gamma_r}\, | 1 \rangle\langle 2 | $,
$ \sigma_s \!\equiv\! \sqrt{\Gamma_s}\, | 1 \rangle\langle 3 | $.
The quantum dot is initialized in the biexciton state: $\rho_{QD}
(0) = | 4\rangle\langle 4|$; in the absence of multiple
excitation-relaxation cycles, the cascade-emission process from such
level results in the generation of two photons. The following time
evolution of $\rho_{QD}$ can be solved analytically (see Appendix
A). Within the AG approach, the photons are delayed in time by a
quantity that depends on their energy and polarization (Fig.
\ref{fig1}). The resulting relations between the input mode $
\sigma_\chi $ and the corresponding output modes $ a_{\alpha i} $
read (up to a common time delay):
\begin{eqnarray}
\label{delay}
a_{H1} (t)  = \sigma_p (t-\delta_p) &,&
a_{H2} (t)  = \sigma_r (t-\delta_r) ,\nonumber\\
a_{V1} (t)  = \sigma_s (t-\delta_s) &,&
a_{V2} (t)  = \sigma_q (t-\delta_q) .
\end{eqnarray}

The quantity of interest here is the degree of entanglement between
the frequency and polarization degrees of freedom of the two-photon
state. This can be computed from their density matrix $\rho_{ph}$
which is derived from the QD dynamics ({\it i.e.}, from $\rho_{QD}$)
through Eqs. (\ref{delay}). In the following, we refer to the basis
$ \{ | H1 , H2 \rangle , | H1 , V2 \rangle ,
     | V1 , H2 \rangle , | V1 , V2 \rangle \} $;
here, the first (second) mode is identified by the central frequency
$ \omega_1\!\equiv\!\omega_p $ ($ \omega_2\!\equiv\!\omega_r $),
which coincides with
$ \omega_s $ ($ \omega_q $)
in the ideal case $\Delta_B=0$.
Within a single cascade decay, and in the absence of non-radiative
relaxation channels, the matrix elements of $\rho_{ph}$ correspond
to the time integrals of second-order correlation functions (see
Appendix B):
\begin{equation}\label{rph}
\langle \alpha 1 , \beta 2 | \rho_{ph} | \gamma 1 , \delta 2 \rangle
=
\int \!\! dt' \int \!\! d\tau' \ G_{\alpha \beta \gamma \delta} ( t' , \tau' ) .
\end{equation}
Here,
$ G_{\gamma \delta \alpha \beta} ( t , \tau )
=
G^{ij}_{\alpha \beta \gamma \delta} ( t , | \tau |) $,
with $ ij = 12 $ for $ \tau > 0 $
and  $ ij = 21 $ for $ \tau < 0 $,
whereas
\begin{eqnarray}\label{g2all}
G^{ij}_{\gamma \delta \alpha \beta} ( t , \tau \!>\! 0)
=
\langle a_{\alpha i}^\dagger (t         )\, a_{\beta  j}^\dagger (t\!+\!\tau) \,
        a_{\gamma j}         (t\!+\!\tau)\, a_{\delta i}         (t         ) \rangle .
\end{eqnarray}
After applying Eqs. (\ref{delay}), the second-order correlation
functions involving the time-shifted ladder operators $\sigma_\chi$
are solved by means of the quantum regression theorem (see Appendix
A). Experimentally, the matrix elements of $ \rho_{ph} $ can be
accessed within a polarization tomography experiment.
\cite{akopian,stevenson06,young06,meirom}

Given the above master equation and initial conditions, only few
elements of the density matrix do not vanish identically (see
Appendixes). These are the diagonal elements $ \rho_{HH} \equiv
\langle H 1 , H 2 | \rho_{ph} | H 1 , H 2 \rangle $ and $ \rho_{VV}
\equiv \langle V 1 , V 2 | \rho_{ph} | V 1 , V 2 \rangle $, and the
off-diagonal one $ \rho_{HV} \equiv \langle H 1 , H 2 | \rho_{ph} |
V 1 , V 2 \rangle $. As a consequence, the two-photon density matrix
reads:
\begin{eqnarray}
\rho_{ph} = \left(
\begin{array}{cccc}
\rho_{HH}   & 0 & 0 & \rho_{VH} \\
     0   & 0 & 0 &      0 \\
     0   & 0 & 0 &      0 \\
\rho_{HV}   & 0 & 0 & \rho_{HH}
\end{array}
\right) . \label{rho_ph}
\end{eqnarray}

The degree of entanglement of the two-photon state can be quantified
by the concurrence ($\mathcal{C}$), whose value ranges from 0 to 1,
going from factorizable to maximally entangled states.
For the above
density matrix, it is easily seen that $ \mathcal{C} (\rho_{ph}) = 2
| \rho_{HV} | $.

\section{Results}\label{results}

\subsection{Two-photon density matrix}

We start by considering the diagonal matrix elements of $\rho_{ph}$,
namely $\rho_{HH}$ and $\rho_{VV}$. The contribution to
$\rho_{HH}^{12}= \langle H 1 , H 2 | \rho_{ph} | H 1 , H 2 \rangle $
corresponding to the ordered detection of photons $ H1$ and $ H2 $
is given by the time integral of the correlation function
\begin{eqnarray}\label{G12HHHH}
    G^{12}_{HHHH} (t',\tau') &\! =\! &
\langle \sigma_p^\dagger (t     )\, \sigma_r^\dagger (t+\tau)\,
        \sigma_r         (t+\tau)\, \sigma_p         (t     )\, \rangle
\nonumber\\
   & = & \Gamma_p\Gamma_r\exp [-(\Gamma_p+\Gamma_q)t-\Gamma_r\tau] ,
\end{eqnarray}
where $ t'\! = \!t+\delta_p$ and $\tau'\! = \!\tau+\delta_r-\delta_p $.
The condition that the $H1$ photon is detected before $H2$ ($\tau'\!>0\!$)
results into a lower bound for the delay in the input modes $\sigma_\chi $:
$ \tau > {\rm max} (0,\delta_r - \delta_p) $.
Analogously, the contribution to $ \langle V 1 , V 2 | \rho_{ph} | V 1 ,
V 2 \rangle $ corresponding to the ordered detection
of photons $ V1$ and $ V2 $, $\rho_{VV}^{12}$, is given by the time integral of the correlation
function:
\begin{eqnarray}\label{G12VVVV}
    G^{12}_{VVVV} (t',\tau') &\! =\! &
\langle \sigma_q^\dagger (t     )\, \sigma_s^\dagger (t+\tau)\,
        \sigma_s         (t+\tau)\, \sigma_q         (t     )\, \rangle
\nonumber\\
   & = & \Gamma_q\Gamma_s\exp [-(\Gamma_p+\Gamma_q)t-\Gamma_s\tau] ,
\end{eqnarray}
where $ t'\! = \!t+\tau+\delta_s$ and $\tau'\! = \!-\tau+\delta_q-\delta_s$.
%
%
 \begin{figure}
 \begin{center}
 \includegraphics[width=\columnwidth]{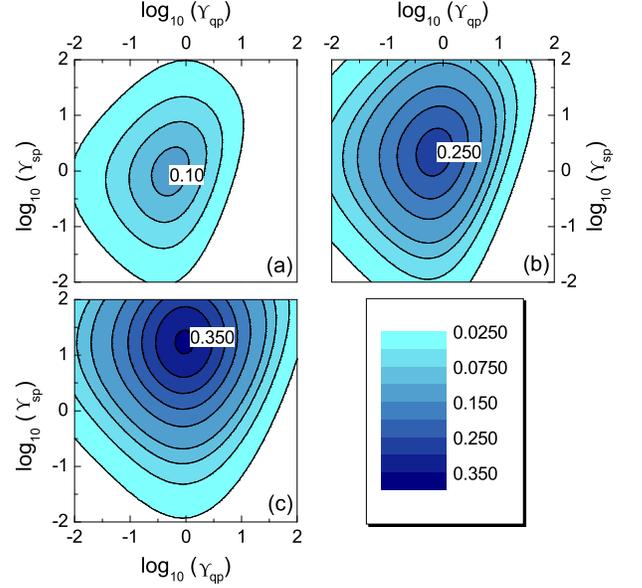}
 \caption{(Color online) Values of $ |\rho_{HV}|= \mathcal{C}^M ({\bf \Gamma }) /2 $ as
 a function of $ \Upsilon_{sp} $ and $ \Upsilon_{qp} $ (where
 $ \Upsilon_{\chi\chi'} \!\! = \!\! \Gamma_\chi / \Gamma_{\chi'} $).
 The value of $ \log (\Upsilon_{rp}) $ has been fixed to $-1.2$ (a),
 0 (b), and $1.2$ (c). }
 \label{fig2}
 \end{center}
 \end{figure}
%
%
Here, the photon order in the output modes ($ a_{V1} $ and $ a_{V2} $)
is inverted with respect to that of the corresponding input modes
($ \sigma_s $ and $ \sigma_q $).
This condition results in an upper bound for the
delay in the emission process: $0<\tau\! < \!\delta_q-\delta_s$.

After time integration in $t$ and $\tau$, the above expressions yeld
the following coincidence probabilities:
\begin{subequations}
\begin{eqnarray}
 \rho_{HH}^{12} & \!=\! &
\Gamma_p\Gamma_r
\int_0^\infty dt \int_{\max(0,\delta_{pr})}^{\infty} d\tau\,
e^{-(\Gamma_p+\Gamma_q)t-\Gamma_r\tau} \nonumber\\
& \!=\! &
\frac{\Gamma_p}{\Gamma_p+\Gamma_q} \{\exp[-\Gamma_r\!\!\times\!\max(0,\delta_{pr})]\},
\\
 \rho_{VV}^{12} & \!=\! &
\Gamma_q\Gamma_s
\int_0^\infty dt \int_{0}^{\delta_{qs}} d\tau\,
e^{-(\Gamma_p+\Gamma_q)t-\Gamma_s\tau} \nonumber\\
& \!=\! &
\frac{\Gamma_q}{\Gamma_p+\Gamma_q} \{1-\exp[-\Gamma_s\!\!\times\!\min(0,\delta_{qs})]\} ,
\end{eqnarray}
\end{subequations}
with $\delta_{\chi\chi'}\!\equiv\!\delta_\chi-\delta_{\chi'}$.
Analogous expressions apply to the case $ij=21$.
After summing up the contributions corresponding to the two cases,
the diagonal elements in the two-photon density matrix take the simple form:
$ \rho_{HH} \! = \! \rho_{HH}^{12} + \rho_{HH}^{21} \! = \! \Gamma_p / (\Gamma_p+\Gamma_q) $
and
$ \rho_{VV}  \! = \! \rho_{VV}^{12} + \rho_{VV}^{21} \! = \! \Gamma_q / (\Gamma_p+\Gamma_q) $.
Even though the two-photon concurrence depends on the off-diagonal terms
of $\rho_{ph}$,
upper limits for $\mathcal{C}$ can already be derived from the diagonal
elements.
In fact, being $ |\rho_{HV}|^2 \le \rho_{HH} \rho_{VV} $, it turns out that
$ \mathcal{C} \le \mathcal{C}_0\equiv\! 2\Upsilon_{pq}^{1/2} / (1+\Upsilon_{pq}) $,
with $ \Upsilon_{\chi\chi'} \!\! \equiv \! \!\Gamma_\chi / \Gamma_{\chi'} $.
Such upper limit, corresponding to the density matrix $\rho_{ph}$
of a pure state, has an absolute maximum of 1 for $\Upsilon_{pq} \!=\! 1$.
The physical interpretation of the above inequality is that,
besides the erasure of the which-path information, a high degree of entanglement
in the two-photon state requires a balanced branching ratio between the $H$
and $V$ decay paths.

The relevant off-diagonal matrix element of $\rho_{ph}$ is given by
the time integral of the correlation functions (Eq. (\ref{rph})):
\begin{eqnarray} \label{g2HHVVp}
G_{HHVV} (t',\tau') \!\!\! &=& \!\!\!
\langle \sigma_p^\dagger (t         ) \sigma_r^\dagger (t\!+\!\tau         )
        \sigma_s         (t\!+\!     \delta_{ps})
        \sigma_q         (t\!+\!\tau-\delta_{qr}) \rangle
\nonumber\\
& = & \exp [A(t,\tau)+iB(\tau)] ,
\end{eqnarray}
where $ t'\! = \!t+\delta_p$ and $\tau'\! = \!\tau+\delta_r-\delta_p $.
The real and imaginary parts of the exponent in the second line are:
\begin{eqnarray}
A(t,\tau) \!\!\!& \equiv &\!\!\! A_0 \!-\!(\Gamma_p\!+\!\Gamma_q)t\!-\!
(\Gamma_p\!+\!\Gamma_q\!+\!\Gamma_r\!-\!\Gamma_s)\tau/2 ,
\\ \label{btau}
B(\tau) \!\!\!& \equiv &\!\!\! B_0\!-\! (E_4\!-\!E_2\!-\!E_3)\tau  ,
\end{eqnarray}
where $ A_0 \!=\!
(\Gamma_p+\Gamma_q)\delta_{qr}/2-\Gamma_s(\delta_{qr}+\delta_{ps})/2
$ and $ B_0 \!=\! (E_4-E_2) \delta_{ps} - E_2 \delta_{qr}$. The
integration intervals result from the requirements that, in Eq.
(\ref{g2HHVVp}), all times in the input modes be positive ($ t+\tau
\!>\! \delta_{qr}, t \! >\! -\delta_{ps}$), and that the biexciton
relaxation takes place before the exciton one
($\tau\!<\!\delta_{ps}\!+\!\delta_{qr}$); otherwise the two-time
expectation value on the right-hand side of Eq. (\ref{g2HHVVp})
vanishes identically. As a consequence, the phase coherence between
the two linearly polarized components of the two-photon state reads:
\begin{equation}\label{gammap}
 \rho_{VH}
=
\int_{\max (0,\delta_{ps})}^{\infty} dt \int_{\delta_{qr} - t}^{\delta_{ps} + \delta_{qr}} d\tau\
e^{A(t,\tau)+iB(\tau)} .
\end{equation}
A finite biexciton binding energy would result in an
oscillating term $e^{iB(\tau)}$,
and therefore in a suppression of $ | \rho_{HV} |$
for
$ |\Delta_B| \!\gtrsim\!
(\Gamma_p + \Gamma_q + \Gamma_r - \Gamma_s)/2 $.
Within the WG strategy, the analogous condition reads:
$ |\delta_{HV}| \!\gtrsim\! (\Gamma_r \!+\! \Gamma_s) / 2$.
If the resonance condition $E_4\!=\!E_2+E_3$
is fulfilled, then $B(\tau)=B_0$, and the (constant)
phase of $G_{HHVV} (t,\tau)$ plays no role.
The concurrence that quantifies the energy-polarization entanglement
of the two-photon state then corresponds to:
\begin{eqnarray}\label{gamma}
\mathcal{C}\!\!\!\!& = &\!\!\!\!
\frac{2\,(\prod_\chi\Gamma_\chi)^{1/2}}{CDE}
\left\{
C {\rm e}^{-[\Gamma_s \delta_{ps} +(\Gamma_p+\Gamma_q+\Gamma_s) \delta_{qr}]/2} +
\right.
\nonumber\\
\!\!\!
& D &\!\!\left.\!\! {\rm e}^{-(\Gamma_s \delta_{ps} \!+\! \Gamma_r \delta_{qr})/2}
\!\! + \!\!
E {\rm e}^{-[\Gamma_r \delta_{qr}\!+\!(\Gamma_p\!+\!\Gamma_q\!+\!\Gamma_r) \delta_{ps} ]/2} \!
\right\}\!\! ,
\end{eqnarray}
where $ C \!\!=\!\! -( \Gamma_p\! + \!\Gamma_q\! + \!\Gamma_r\! -
\!\Gamma_s ) / 2 $, $ E \!\!=\!\! -( \Gamma_p\! + \!\Gamma_q\! -
\!\Gamma_r\! + \!\Gamma_s ) / 2 $, and $ D \!\!=\!\! \Gamma_p\! +
\!\Gamma_q $. We note that $\mathcal{C}$ only depends on
$\delta_{ps}$ and $ \delta_{qr}$, and not on the four delays
$\delta_\chi$ independently. This is consistent with the intuition
that the degree of entanglement depends on the extent to which the
delays make the $H1$ ($H2$) mode indistinguishable from $V1$ ($V2$)
in the time domain.

\subsection{Parameter optimization}

Given the analytic expression of the concurrence,
$\mathcal{C} (\delta_{ps},\delta_{qr}; {\bf \Gamma})$,
we first maximize it with respect to the delays, as a function of the
relaxation rates.
The optimum values of the delays are denoted with
$ \delta_{ps}^M ({\bf \Gamma}) $ and $ \delta_{qr}^M ({\bf \Gamma})$,
being
$ {\bf \Gamma} \!=\! (\Gamma_p,\Gamma_q,\Gamma_r,\Gamma_s) $.
The corresponding concurrence is
\begin{equation}
\mathcal{C}^M ({\bf \Gamma})
\equiv
\mathcal{C} [\delta_{ps}^M ({\bf \Gamma}),\delta_{qr}^M ({\bf \Gamma}); {\bf \Gamma}] .
\end{equation}
In a second step, we maximize $ \mathcal{C}^M ({\bf \Gamma}) $ with respect to
the relaxation rate, thus identifying the absolute maximum of $\mathcal{C}$, namely
$ \mathcal{C}_{opt} \equiv \mathcal{C}^M ({\bf \Gamma}_{opt} )$.

The values of the relative delays that maximize $\mathcal{C}$
satisfy the conditions:
$ \partial \mathcal{C} / \partial \delta_{ps} = 0 $
and
$ \partial \mathcal{C} / \partial \delta_{qr} = 0 $.
Their expressions read:
\begin{subequations}
\begin{eqnarray}\label{delaysa}
\delta_{ps}^M ({\bf \Gamma}) \!\! &=& \!\! \frac{2}{\sum_\chi\Gamma_\chi-2\Gamma_s}
\ln \left( \frac{\sum_\chi\Gamma_\chi}{2\Gamma_s} \right) ,
\\\label{delaysb}
\delta_{qr}^M ({\bf \Gamma}) \!\! &=& \!\! \frac{2}{\sum_\chi\Gamma_\chi-2\Gamma_r}
\ln \left( \frac{\sum_\chi\Gamma_\chi}{2\Gamma_r} \right) .
\end{eqnarray}
\end{subequations}

%
%
 \begin{figure}
 \begin{center}
 \includegraphics[width=\columnwidth]{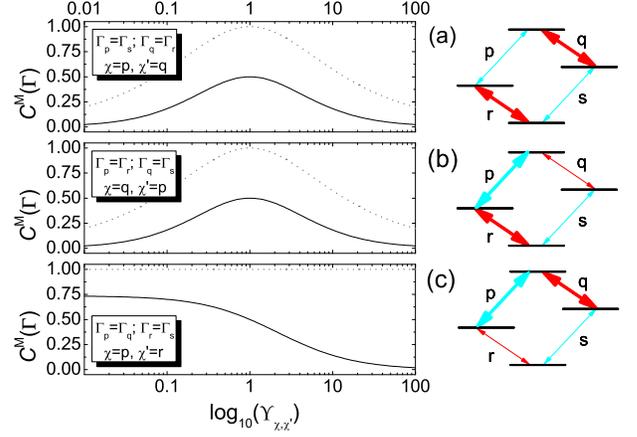}
 \caption{(Color online) The solid black lines give the
values of the concurrence for optimized delays, $\mathcal{C}^{M}$,
as a function of the ratio $\Upsilon_{\chi,\chi'} = \Gamma_\chi /
\Gamma_{\chi'}$ and for different constraints. (a) $ \Gamma_s =
\Gamma_p $ and $ \Gamma_r = \Gamma_q $, with $(\chi , \chi') =
(q,p)$; (b) $ \Gamma_s = \Gamma_q $ and $ \Gamma_r = \Gamma_p $,
with $(\chi , \chi') = (p,q)$; (c) $ \Gamma_q = \Gamma_p $ and $
\Gamma_s = \Gamma_r $, with $(\chi , \chi') = (p,r)$. The dashed
lines are the values of the upper limits for the concurrence
$\mathcal{C}_0$, corresponding to a complete cancelation of the
which-path information. Photon emission rates that are set equal are
denoted by arrows of equal thickness in the level schemes on the
right-hand side of each plot.}
 \label{fig3}
 \end{center}
 \end{figure}
%
%
In the general case, one can substitute the above equations in
Eq.~(\ref{gamma}), thus obtaining $ \mathcal{C}^{M} ({\bf\Gamma })
$. In order to maximize the concurrence, we look for the values of
$\Gamma_\chi$ that satisfy the equations: $\partial \mathcal{C}^{M}
/ \partial \Gamma_\chi = 0 $. No value of ${\bf\Gamma }$
simultaneously fulfils these conditions. Therefore, we numerically
compute $ \mathcal{C}^M ({\bf \Gamma}) $ for a wide range of
relaxation-rates values: $ \Upsilon_m < \Upsilon_{qp} ,
\Upsilon_{rp} , \Upsilon_{sp} < \Upsilon_M $. The global maximum of
the concurrence corresponds to $ \Gamma_p \!=\! \Gamma_q $ and $
\Gamma_r \! = \! \Gamma_s $, with the latter rates much larger than
the former ones:
\begin{equation}
\mathcal{C}_{opt} =
\lim_{\Upsilon_{pr} \rightarrow 0}
\mathcal{C}^M ({\bf \Gamma} ) \simeq 0.736.
\end{equation}
In order to gain some further understanding, we plot the dependence of
$ \mathcal{C} $ on $\Gamma_q$ and $\Gamma_s$, for three different
values of $\Gamma_r$, all in units of $\Gamma_p$ (see Fig.~\ref{fig2}).
On average, $ \mathcal{C} $ increases with increasing values of
$ \Gamma_r / \Gamma_p $ [from panel (a) to (c)].
For $ \Gamma_r \!<\! \Gamma_p $ [panel (a)], the maximum is localized close
to the point $ \Upsilon_{sp}\! =\! \Upsilon_{qp}\! =\! 1 $.
The optimum delays in this case [$\log ( \Upsilon_{rp} ) \! =\! -1.2$] are
no longer identical. In particular, they are given by:
$\delta_{ps}^M \simeq 0.80/\Gamma_p$
and
$\delta_{qr}^M \simeq 2.2 /\Gamma_p$.
If the $X_H$ exciton relaxation rate is larger than the biexciton one
[see panel (c), where $\log ( \Upsilon_{rp} ) \! =\! -1.2$],
the maximum of $\mathcal{C}$ is localized close
to the point $ \Upsilon_{sp}\! =\! 1.2 $, $\Upsilon_{qp}\! =\! 1 $.
The corresponding delays are reduced to:
$\delta_{ps}^M \! = \! \delta_{qr}^M \simeq 0.061 / \Gamma_p $.
The above examples show that the best values of the delays strongly
depend on the relaxation rates.
As to the dependence of $\mathcal{C}$ on ${\bf\Gamma}$, the above plots
suggest that the two biexciton (exciton) relaxation rates should coincide
(are correlated).

As already mentioned, the exciton and biexciton relaxation rates can, to
some extent, be
engineered in the growth process.~\cite{wimmer}
A further tuning can be achieved by coupling the dot with an optical
microcavity, through the Purcell effect.~\cite{pelton}
In the following, we further consider the dependence of the concurrence
optimized with respect to the delays,
$ \mathcal{C}^M ({\bf \Gamma}) $,
after introducing specific relations between the parameters $\Gamma_\chi$.
These will either be realizable by suitable dot-cavity couplings,
starting from a situation where $\Gamma_\chi \simeq \Gamma $ [cases
$(i)$ and $(ii)$], or correspond to a region of specific relevance for
the maximization of $\mathcal{C}$ $(iii)$.

$(i)$ In the first case, the two decay processes with equal frequencies
share the same value of the emission rates:
$ \Gamma_p \!\! = \!\! \Gamma_s $
and
$ \Gamma_q \!\! = \!\! \Gamma_r $.
This condition might be fulfilled by coupling the QD with a suitable
microcavity. In particular, the MC should possess a mode doubly degenerate
with respect to polarization, and in resonance with two QD transitions
(e.g., $p$ and $s$), while sufficiently off-resonance with the remaining
two. The Purcell effect would then result in an effective,
frequency-selective enhancement of the emission rates $\Gamma_\chi$.
The optimum values of the delays reduce to:
$ \delta_{ps}^M = (1/\Gamma_q) \ln ( \Gamma_q / \Gamma_p + 1 ) $
and
$ \delta_{qr}^M = (1/\Gamma_p) \ln ( \Gamma_p / \Gamma_q + 1 ) $.
After substituting these expressions in Eq. (\ref{gamma}), we
obtain [black curve, Fig. \ref{fig3} (a)]:
\begin{equation}\label{gammaM}
\mathcal{C}^M ({\bf \Gamma} |
\Upsilon_{ps} , \Upsilon_{qr} \!\! = \!\! 1 ) =
\frac{4\Upsilon_{qp}^{1+\Upsilon_{qp}/2}}{(1+\Upsilon_{qp})^{2+\Upsilon_{qp}/2+1/(2\Upsilon_{qp})}} .
\end{equation}
The extrema of such function are identified by the zeros of its
derivative with respect to $ \Upsilon_{qp} $. The constrained
maximum of the concurrence is: $ \mathcal{C}^M ({\bf \Gamma} |
\Upsilon_{ps} , \Upsilon_{qr} \!\! = \!\! 1 ) $, which is $
\mathcal{C}_{opt} ({\bf \Gamma} | \Upsilon_{ps} , \Upsilon_{ps} \!\!
= \!\! 1 ) \! = \! 1 /2$, for $\Upsilon_{qp} \!\!=\!\! = 1/2 $. This
value corresponds to half of upper limit for the concurrence,
$\mathcal{C}_0$, in the case $ \rho_{HH} \!\! = \!\! \rho_{VV} \!\!
= \!\! 1/2 $ (gray curve).

$(ii)$ In the second case, the relaxation rates depend only on
polarization: $ \Gamma_p \!\! = \!\! \Gamma_r $ and $ \Gamma_q \!\!
= \!\! \Gamma_s $. Such situation can be induced by a cavity with a
linearly polarized mode, sufficiently broadened in frequency so as
to couple to both the QD transitions of a given linear polarization
($H$ or $V$), while remaining uncoupled with the other one. Given
these constraints, the delays become: $ \delta_{ps}^M = (1/\Gamma_p)
\ln ( \Gamma_p / \Gamma_q + 1 ) $ and $ \delta_{qr}^M = (1/\Gamma_q)
\ln ( \Gamma_q / \Gamma_p + 1 ) $. This results in an expression for
the concurrence that coincides with that of above case $(i)$ [Fig.
\ref{fig3} (b)]. In fact, from Eqs. (\ref{gamma}), (\ref{delaysa}),
(\ref{delaysb}) one can see that $ \mathcal{C}_{M}^{(\Gamma )} $ is
invariant under the simultaneous exchange $\Gamma_p
\longleftrightarrow \Gamma_q$, and $\Gamma_r \longleftrightarrow
\Gamma_s$: $ \mathcal{C}^M ( \Gamma_q , \Gamma_p , \Gamma_s ,
\Gamma_r ) = \mathcal{C}^M ( \Gamma_p , \Gamma_q , \Gamma_r ,
\Gamma_s ) $ . We incidentally note that this is not true in general
for arbitrary values of the delays, {\it i.e.} if $ ( \delta_{ps} ,
\delta_{qr}   ) \neq
  ( \delta_{ps}^M , \delta_{qr}^M ) $.
For $\Upsilon_{pq}\neq 1$, the upper limit $\mathcal{C}_0<1$ (gray
curve): thus, the two-photon concurrence cannot attain its maximum
value even for a complete cancelation of the which-path information,
simply because of the asymmetric branching ratio between the $H$ and
$V$ paths. The difference between $\mathcal{C}$ and $\mathcal{C}_0$,
instead, can be ascribed to the distinguishability between the
wavepackets relevant to the two polarizations. Therefore neither the
energy [panel (a)], nor the polarization-selective tuning [panel
(b)] of the QD photon-emission rates allow to achieve high values of
the concurrence, and specifically to cancel the which-path
information required in the AG scheme.

$(iii)$ In the third case, the biexciton and the exciton relaxation rates are
independent on the polarization:
$ \Gamma_p \!\! = \!\! \Gamma_q $ and $ \Gamma_r \!\! = \!\! \Gamma_s $.
The optimum delays then read:
$\delta_{qr}^M \!=\! \delta_{ps}^M \!=\! (1 / \Gamma_p) \ln (1+\Gamma_p/\Gamma_r) $,
resulting in (black curve)
\begin{equation}
\mathcal{C}^M ({\bf \Gamma} | \Upsilon_{pq} , \Upsilon_{rs} \!\! = \!\! 1 )
= 2 (\Upsilon_{pr} + 1)^{-(1+1/\Upsilon_{pr})} .
\end{equation}
The above expression is a decreasing function of $ \Upsilon_{pr} $;
it tends to $ \mathcal{C}_M \!=\! 2 / e $ for $ \Upsilon_{pr}
\rightarrow 0 $, {\it i.e.}, in the limit of biexciton relaxation
much slower than the exciton one. This limiting value coincides with
the global maximum that we find for the unconstrained case.
Therefore, as already reported above and in Fig. \ref{fig2}, the
most favorable region in the parameter space ${\bf \Gamma}$
corresponds to the biexciton and exciton relaxation rates being
independent on polarization, with the former ones much smaller than
the latter ones. Unfortunately, the present case seems to be the
least feasible. In fact, within the AG approach (where $\omega_p
\!\! = \!\! \omega_s \!\!\neq\!\! \omega_q \!\! = \!\! \omega_r $),
the transitions $ B \!\rightarrow\! X_{H/V} $ cannot be resolved
from the $ X_{H/V} \!\rightarrow\! G $ ones, neither spectrally nor
through polarization. This impedes to optimize the relaxation rates
through the Purcell effect induced by dot-cavity coupling, and
forces to rely on the engineering of the QD oscillator strengths
alone. We finally note that the case $ ( \Gamma_p , \Gamma_r ) = (
\Gamma_q , \Gamma_s ) $ is the one considered throughout Ref.
\onlinecite{avron}. There, analogous conclusions are drawn with
respect to the dependence of the two-photon entanglement on the
ratio $ \Gamma_p / \Gamma_r $. However, we find that the optimized
delays
 differ from those
suggested by Avron and coworkers, apart from the limiting case
$ \Gamma_r \gg \Gamma_p $, where
$\delta_{qr}^M \!=\! \delta_{ps}^M \rightarrow 1/\Gamma_r$.

\section{Conclusions} \label{conclusions}

We have theoretically investigated the generation of
energy-polarization entanglement of two-photons emitted by the
cascade decay QD within the AG approach. As in the case of the WG
scheme, the two-photon entanglement is limited by dephasing and
imperfect dot initialization. Unlike that case, the concurrence
$\mathcal{C}$ is also limited by the opposite emission order of the
$\omega_1$ and $\omega_2$ photons along the $H$ and $V$ paths (see
Fig. \ref{fig1}). A unitary erasure of the which-path information
can be performed by time reordering. \cite{avron} Here, we have
analytically computed $\mathcal{C}$ as a function of the parameter
that determine the time-reordering ({\it i.e.}, the delays
$\delta_\chi$) and characterize the two-photon source ({\it i.e.},
the photon-emission rates $\Gamma_\chi$). We have maximized
$\mathcal{C}$ with respect to $\delta_\chi$, as a function of
$\Gamma_\chi$, thus optimizing the erasure process for an arbitrary
source. The optimized concurrence $\mathcal{C} ({\bf \Gamma})$ has
then been maximized with respect to the emission rates, thus
providing indications on the desirable source engineering. We find
that both the energy and polarization selective enhancement of the
emission rates, that could be induced by suitably coupling the QD
with an optical microcavity (Purcell effect), are of limited
usefulness. In fact, the maximum value $\mathcal{C}=0.5$ corresponds
to identical rates ($\Gamma_\chi = \Gamma$). On the other hand, the
absolute maximum of the concurrence, $\mathcal{C}=2/e\simeq 0.736$,
corresponds to biexciton and exciton relaxation rates independent on
polarization, with the former ones much smaller than the latter ones
($ \Gamma_p = \Gamma_q \ll \Gamma_r = \Gamma_s $). However, within
the AG approach (where $\omega_p \!\! = \!\! \omega_s \!\!\neq\!\!
\omega_q \!\! = \!\! \omega_r $), the transitions $ B
\!\rightarrow\! X_{H/V} $ cannot be resolved from the $ X_{H/V}
\!\rightarrow\! G $ ones, neither spectrally nor through
polarization. This impedes to access the above regime through the
Purcell effect.

This work has been partly supported by Italian MIUR under FIRB
Contract No. RBIN01EY74; by the spanish MEC under the contracts
QOIT-Consolider-CSD2006-0019, MAT2005-01388, NAN2004-09109-C04-4;
and by CAM under contract S-0505/ESP-0200.

\section*{APPENDIX A: DYNAMICS AND QUANTUM-REGRESSION THEOREM}

Given the initial conditions
$ \rho_{QD} (0) = | 4\rangle\langle 4| $,
the time evolution of the QD density matrix induced by the Liouvillian
$ \dot{\rho}_{QD} (t) = i[\rho_{QD},H] + \sum_\chi \mathcal{L}_\chi\rho_{QD}
\equiv \mathcal{L}_{QD} (t)\rho_{QD} (0) $
is the following:
\begin{eqnarray}
\langle 4 | \rho_{QD} | 4 \rangle &=& e^{-(\Gamma_p+\Gamma_q)t} ,
\nonumber\\
\langle 3 | \rho_{QD} | 3 \rangle &=& \frac{\Gamma_q}{\Gamma_s-\Gamma_p-\Gamma_q}
\left[ e^{-\Gamma_s t} - e^{-(\Gamma_p+\Gamma_q)t} \right] ,
\nonumber\\
\langle 2 | \rho_{QD} | 2 \rangle &=& \frac{\Gamma_p}{\Gamma_r-\Gamma_p-\Gamma_q}
\left[ e^{-\Gamma_r t} - e^{-(\Gamma_p+\Gamma_q)t} \right] ,
\nonumber\\
\langle 1 | \rho_{QD} | 1 \rangle &=& 1-\sum_{k=2}^4 \langle k | \rho_{QD} | k \rangle .
\end{eqnarray}
The Liouvillian $\mathcal{L}_{QD}$ doesn't couple the diagonal terms
of $\rho_{QD}$ with the off-diagonal ones. Therefore, for the above initial
conditions, these are identically zero.

In order to compute the two-time expectation values
$G_{\alpha\alpha\alpha\alpha}^{ij}$ ($\alpha = H,V$),
we apply the quantum regression theorem (see, e.g.,
Ref. \onlinecite{scully}).
This states that if, for some operator $O$, the time dependence of the
expectation value is given by
\begin{equation} \label{app1}
\langle O(t+\tau) \rangle
=
\sum_j a_j ( \tau ) \langle O_j(t) \rangle ,
\end{equation}
then
\begin{equation} \label{app2}
\langle A(t) O(t+\tau) B(t) \rangle
=
\sum_j a_j ( \tau ) \langle A(t) O_j(t) B(t) \rangle .
\end{equation}
In the case of $G_{\alpha\alpha\alpha\alpha}^{ij}$, after performing
the substitutions reported in Eqs. (\ref{delay}), the above
operators are: $ O = \sigma^\dagger_\chi \sigma_\chi $, $ A =
\sigma^\dagger_{\chi'} $, $ B = \sigma_{\chi'} $, with $ (\chi ,
\chi' ) = (r,p) $ for $ \alpha = H$ and $ (\chi , \chi' ) = (s,q) $
for $ \alpha = V$. Besides, $ O_j = | k(j) \rangle\langle l(j) | $
(with $k,l = 1,2,3,4$ denoting the QD state); their expectation
values $ \langle O_j \rangle = \rho_{l,k} $ correspond to the
elements of the quantum-dot density matrix.

In Eq. (\ref{app1}), the expectation values $ \langle O_j (t)
\rangle $ give the initial conditions of dot state. Therefore, the
two-time expectation value in Eq. (\ref{app2}) corresponds to the
single-time expectation value of $O$, for initial conditions $
\rho_{m,n}' (t) = \langle A(t) O_j (t) B(t) \rangle = \langle m|A(t)
| k(j) \rangle\langle l(j) | B(t) |n\rangle $. This provides an
intuitive explanation of why most of the matrix elements of
$\rho_{ph}$ vanish identically. The element $ \langle
H1,V2|\rho_{ph}| H1,V2\rangle $, for example, corresponds to the
expectation value of $\sigma_q^\dagger \sigma_q = |4\rangle\langle
4|$, for a system initialized in the state $ \rho_{QD}' = \sigma_p
\rho_{QD} (t) \sigma^\dagger_p \propto |2\rangle\langle 2|$. Such
expectation value is zero at any $ \tau $ for $
\mathcal{L}_{QD}(\tau) |2\rangle\langle 2| $: if the dot state is
initialized to the exciton level $X_H$ at time $t$, it will never be
found in the biexciton state at any time $t+\tau$.

The procedure is the analogous for the calculation of the
correlation functions $ G_{HHVV} (t,\tau)$. After performing the
substitutions reported in Eqs. (\ref{delay}), however, this becomes
a four-time expectation value: $ G_{HHVV}^{ij} = \langle
\sigma^\dagger_p (t_1)
          \sigma^\dagger_r (t_2)
          \sigma_s (t_3)
          \sigma_q (t_4) \rangle $
(where, e.g., $ t_1 < t_4 < t_2 < t_3 $).
This is computed by applying three times the quantum regression
theorem, with $ O = \sigma_s $
\begin{eqnarray}
& & A^{(1)} \!=\! \sigma^\dagger_p , \
A^{(2)} \!=\! {I}, \
A^{(3)} \!=\! \sigma^\dagger_r
\nonumber\\
& & B^{(1)} \!=\! {I}, \
B^{(2)} \!=\! \sigma_q , \
B^{(3)} \!=\! {I}
\end{eqnarray}
($I$ is the identity operator).

\section*{APPENDIX B: TWO-PHOTON DENSITY MATRIX}

In the present conditions, where the QD is initialized into the
biexciton state $ |4\rangle $ and undergoes a single cascade decay,
the probability that of detecting a photon in the $\alpha 1$ mode
and a photon in the $\alpha 2$ mode is given by the time integrals
of the second-order correlation functions
$ G^{12}_{\alpha\alpha\alpha\alpha} (t,\tau) $.
The identification of such integrals with the diagonal elements of
the two-photon density matrix,
$\langle\alpha 1,\alpha 2 |\rho_{ph}|\alpha 1 , \alpha 2\rangle$,
results from the fact these have the same physical interpretation.

For the off-diagonal terms the validity of Eq. (\ref{rph}) is less
intuitive. Such corresponds results from the two following points. $
(i) $ The correlation functions $ G_{\alpha\beta\gamma\delta}^{ij}
(t,\tau) $ and the two-photon matrix elements $\langle \gamma 1,
\delta 2 |\rho_{ph}| \alpha 1, \beta 2\rangle $ transform according
to the same equations under the change of polarization basis in the
1 and 2 modes. $(ii)$ The off-diagonal correlation-functions, such
as $ G^{12}_{HHVV} (t,\tau) $, can be expressed as linear
combinations of diagonal ones, namely $
G^{12}_{\alpha\beta\beta\alpha} (t,\tau) $, where $ \alpha $ and $
\beta $ vary over 4 independent photon polarizations (including $H$
and $V$). These are, in fact, the relations that are exploited in
polarization quantum tomography. \cite{tomography} Therefore,
\begin{eqnarray}
\mathcal{G}_{HHVV} &=&
\sum_{\alpha\beta} \Lambda^{HV}_{\alpha\beta}
\int G^{12}_{\alpha\beta\beta\alpha} (t,\tau)\, dt \, d\tau
\nonumber\\
&=&\sum_{\alpha\beta} \Lambda^{HV}_{\alpha\beta}
\langle \alpha 1, \beta 2 |\rho_{ph}| \alpha 1, \beta 2\rangle
\nonumber\\
&=&\langle V 1, V 2 |\rho_{ph}| H 1, H 2\rangle .
\end{eqnarray}
where $ \mathcal{G}_{HHVV} \equiv \int G^{12}_{HHVV} (t,\tau)\, dt
\, d\tau $ and $\Lambda^{HV}$ being the transformation changing the
polarization basis.

\bibliography{paper}

\bibliographystyle{apsrev}

\end{document}